\documentclass{jaa}

\usepackage{natbib}
\usepackage{graphicx}

\begin{document}\sloppy

\title{TIRCAM2 Camera Interface on the Side port of the 3.6 meter Devasthal Optical Telescope.}


\author{Shailesh B. Bhagat\textsuperscript{1}, Milind B. Naik\textsuperscript{1,*}, Satheesha S. Poojary\textsuperscript{1}, Harshit Shah\textsuperscript{1}, Rajesh B. Jadhav\textsuperscript{1}, Balu G. Bagade\textsuperscript{1}, Savio L. D'Costa\textsuperscript{1}, B. Krishna Reddy\textsuperscript{2}, Nadish Nanjappa\textsuperscript{2}, Tarun Bangia\textsuperscript{2}, Devendra K. Ojha\textsuperscript{1}, Saurabh Sharma\textsuperscript{2} and Koshvendra Singh\textsuperscript{1}}
\affilOne{\textsuperscript{1}Department of Astronomy and Astrophysics, Tata Institute of Fundamental Research (TIFR), Mumbai 400005, India.\\}
\affilTwo{\textsuperscript{2}Aryabhatta Research Institute of Observational Sciences (ARIES), Manora Peak, Nainital 263001, India.}


\twocolumn[{

\maketitle

\corres{mbnaik@tifr.res.in}

\msinfo{1 July 2021}{1 July 2021}

\begin{abstract}
The TIFR Near Infrared Imaging Camera-II (TIRCAM2) is being used at the 3.6 m  Devasthal Optical Telescope (DOT) operated by Aryabhatta Research Institute of Observational Sciences (ARIES), Nainital, Uttarakhand, India. Earlier, the TIRCAM2 was used at the main port of the DOT on time shared basis. It has now been installed at the side port of the telescope. Side port installation allows near simultaneous observations with the main port instrument as well as longer operating periods. Thus, the TIRCAM2 serves the astronomical community for a variety of observations ranging from lunar occultations, transient events and normal scheduled observations. 
\end{abstract}


}]


\doinum{12.3456/s78910-011-012-3}
\artcitid{\#\#\#\#}
\volnum{000}
\year{0000}
\pgrange{1--}
\pgrange{1-5}
\setcounter{page}{1}
\setcounter{page}{1}
\lp{1}

\section{Introduction}

The 3.6 meter Devasthal Optical Telescope (DOT) has been built by the Aryabhatta Research Institute of Observational Sciences (ARIES) in collaboration with the Belgian firm, Advanced Mechanical and Optical System \citep[AMOS, see Figure 1 (left),][]{2019CSci..117..365S}. The DOT is located at the Devasthal observatory site near Nainital (Uttarakhand district) at an altitude of 2450 m above mean sea level. The DOT is currently the largest single mirror telescope in India.  The operational wavebands of the DOT run from 0.35 $\mu$m to 5 $\mu$m. Some of the specifications of the DOT are given in Figure 2.

\section{TIRCAM2 - an NIR imaging camera at DOT}

The TIRCAM2 has imaging capability from 1 $\mu$m to 3.7 $\mu$m \citep{2012BASI...40..531N,2018JAI.....750003B}. Figure 1 (right) shows the TIRCAM2 instrument. The camera has an InSb focal plane array with 512 x 512 pixels. It operates at a temperature of 35 K, cooled by a closed cycle Helium cryo-cooler. Many successful science observations have been done with this camera from the main (axial) port of the DOT. For example,
\citet{2018JAI.....750003B} describe imaging capabilities and results of TIRCAM2 in 
J, H, K and narrow-band L (nbL) bands. \citet{2020JApA...41...27A} present imaging capabilities in polycyclic aromatic hydrocarbon and nbL bands. \citep{2020MNRAS.498.2263R}
present milliarcsecond resolution results on cool giants and binary stars from lunar occultations at Devashtal. \citet{2020MNRAS.498.2309S} describe a detailed analysis of the Cz3 open cluster using deep NIR observations taken from TIRCAM2.
After commissioning of the TIFR-ARIES Near Infrared Spectrometer (TANSPEC), an NIR imager and spectrometer at DOT, it was decided to shift the TIRCAM2 camera to the side port of the DOT, thus becoming the first instrument at the side port of the DOT. This required designing a new mechanical structure to interface the TIRCAM2 considering many technical aspects such as optical alignments, mechanical envelop, TIRCAM2's requirements of cryogenic cooling setup, etc.

\begin{figure}[!t]
\includegraphics[width=1\columnwidth]{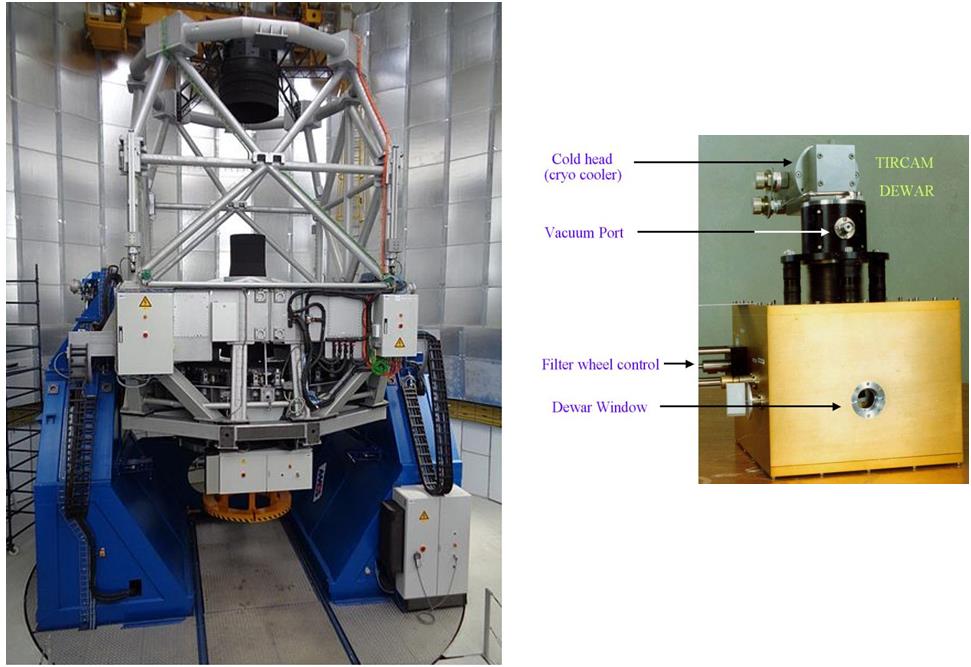}
\caption{Left: The 3.6 meter Devasthal Optical Telescope. Right: The TIRCAM2 imaging camera.}
\label{fig1}
\end{figure}

\begin{figure}[!t]
\includegraphics[width=0.8\columnwidth]{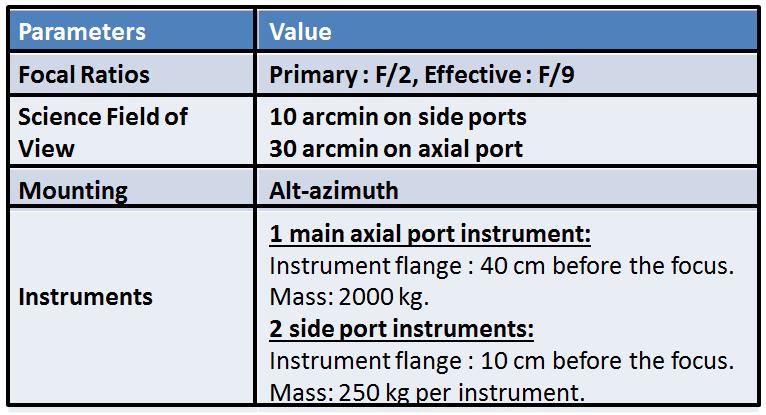}
\caption{Some of the specifications of the DOT.}
\label{fig2}
\end{figure}

\section{Instrument Envelop for the side port}

The DOT provides three cassegrain ports for mounting the instruments, one is the main axial port and other two are the side ports. Figures 3 and 4 show the envelops for designing and developing instruments for the main axial port and the side ports of the DOT. 
Figure 3 shows the front view of the main axial port and both side ports and 
Figure 4 shows the top view of the side port instrument envelops for DOT.
TIRCAM2 main port envelop was 1020 mm (H) $\times$ 612 mm (W) $\times$ 712 mm (L) whereas TIRCAM2 side port envelop is 862 mm (H) $\times$ 360 mm (W) $\times$ 800 mm (L). The side port instruments can have a weight of 250 kg each, with a center of gravity at a position of 620 mm away from the instrument interface plate. The focal plane is available at 100 mm from the interface plate and 620 mm from the fold mirror. Actual photograph of one of the side ports with and without dummy cover is shown in Figure 5.

\begin{figure}[!h]
\includegraphics[width=1\columnwidth]{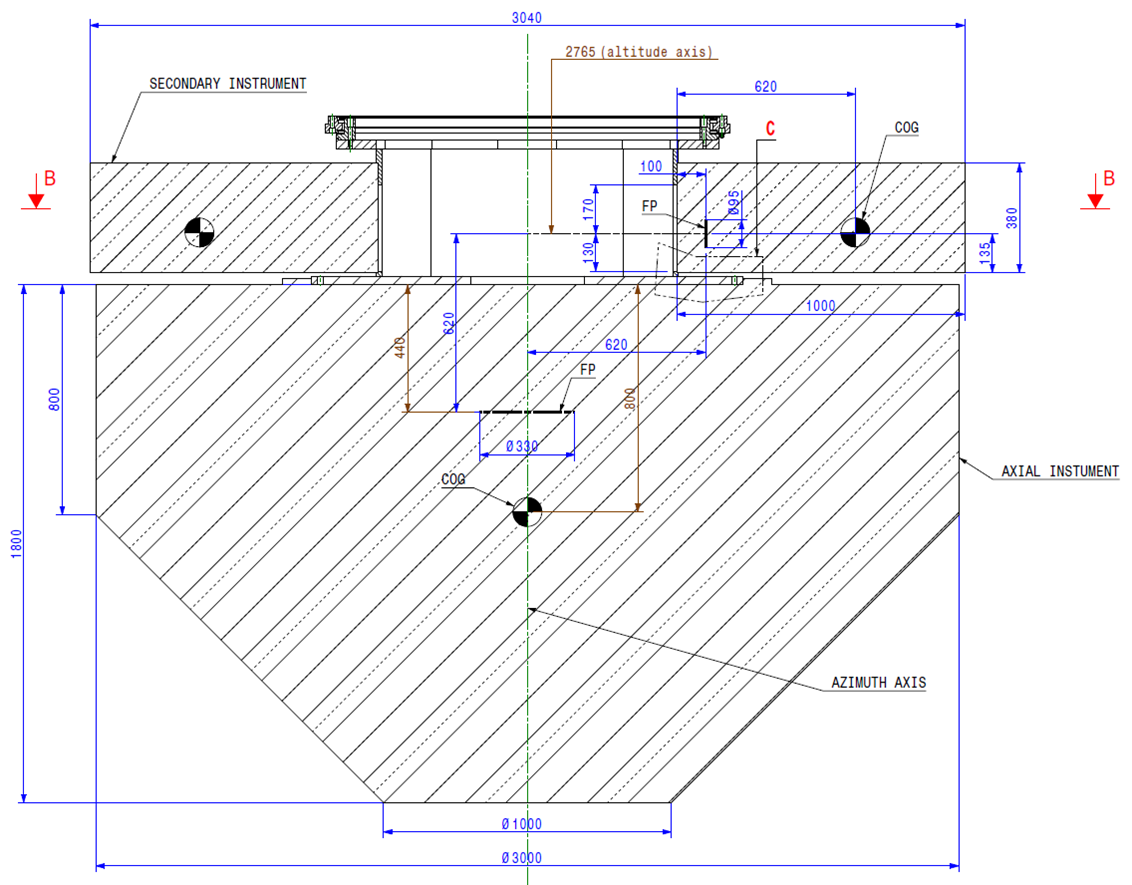}
\caption {Front view of the focal planes showing the main axial port and both side 
ports' instrument envelops for DOT.} 
\label{fig3}
\end{figure}

\begin{figure}[!h]
\includegraphics[width=1\columnwidth]{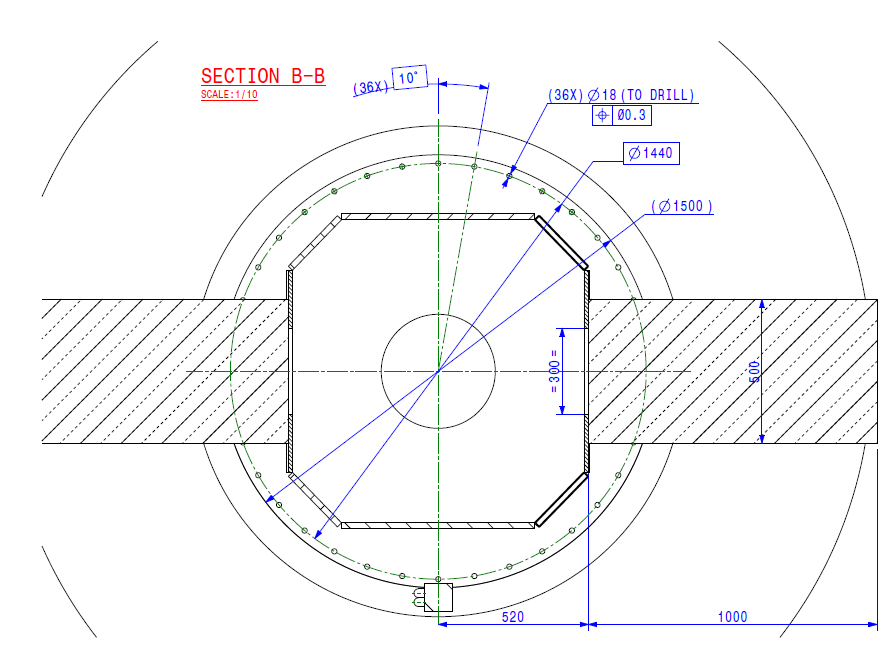}
\caption {Top view of the side port instrument envelops for DOT.} 
\label{fig4}
\end{figure}

\begin{figure}[!h]
\includegraphics[width=1\columnwidth]{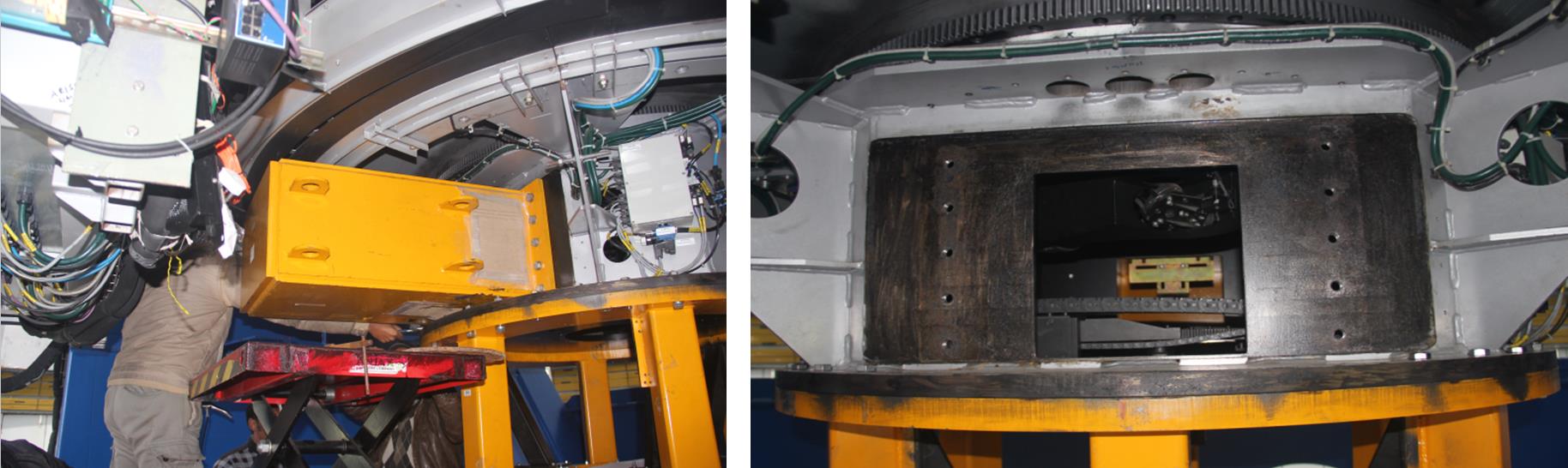}
\caption{The DOT side port with dummy cover (left) and without dummy cover (right).}\label{fig5}
\end{figure}

\section{Mechanical structure design}
Figure 6 shows the Zemax raytrace of the DOT telescope with TIRCAM2 on the side port of DOT. \citet{2012BASI...40..531N} have presented a detailed 
schematic and Zemax raytrace of TIRCAM2 optics. 
The telescope side port focal plane is 100 mm away from the instrument interface plate of the side port and 620 mm away from the fold mirror (see Figure 3). The TIRCAM2 window is at a distance of 587 mm and 67 mm from the fold mirror and the interface plate, respectively. The aperture plane of the TIRCAM2 is 33 mm behind the front surface of the window. AutoCAD and SOLIDWORKS models of the mounting are shown in Figures 7 and 8. Figure 9 shows the AutoCAD drawing and SOLIDWORKS model view of the TIRCAM2 camera with instrument rack mounted on the side port of the telescope. 

\begin{figure}[!h]
\includegraphics[width=1\columnwidth]{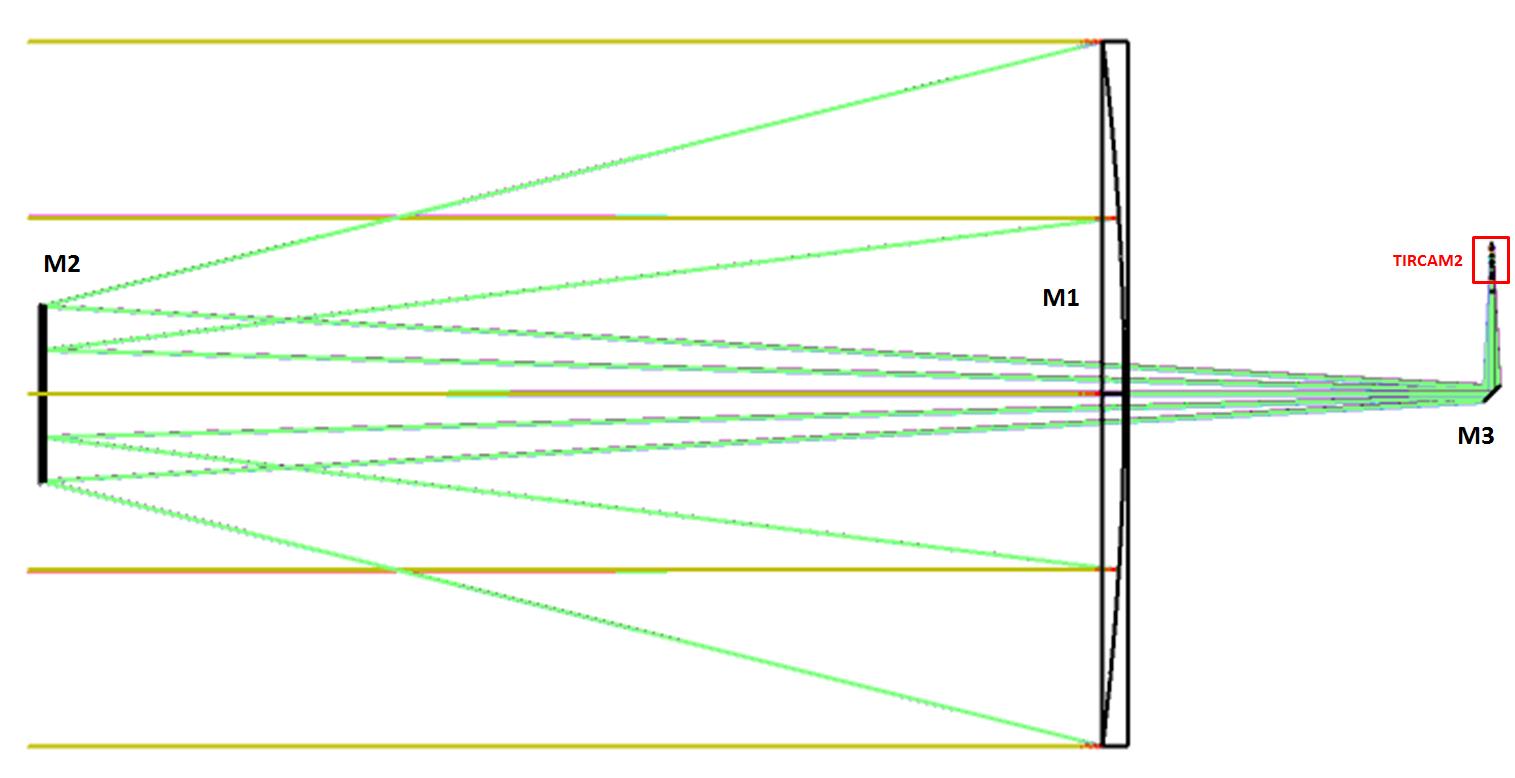}
\caption{Raytrace of the DOT with the TIRCAM2 camera on the side port.}
\label{fig6}
\end{figure}

\begin{figure}[!h]
\includegraphics[width=1\columnwidth]{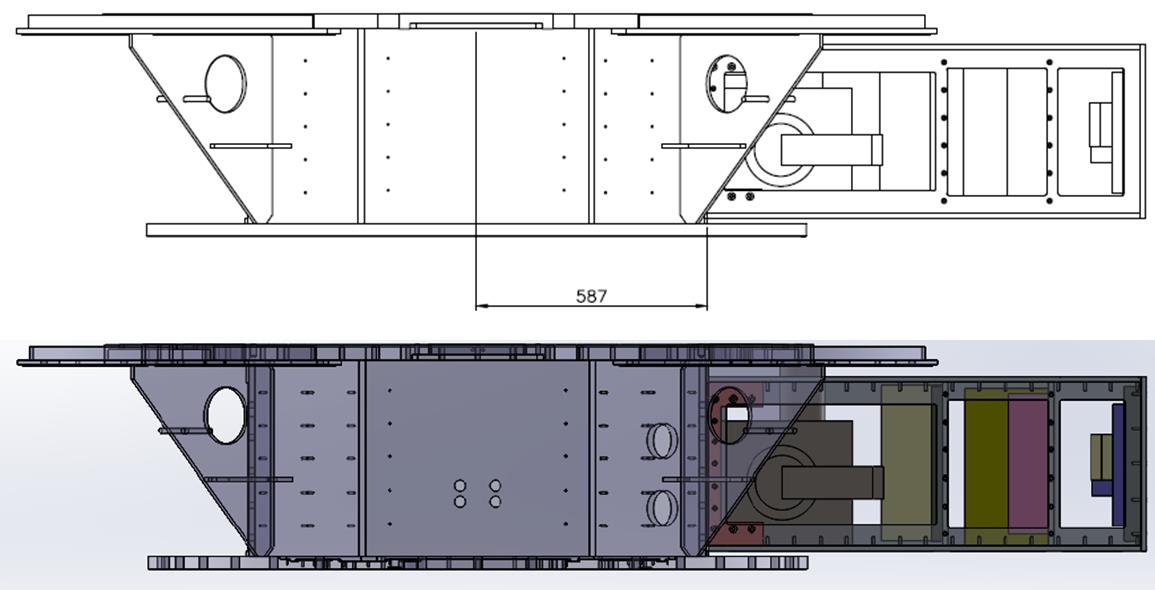}
\caption{AutoCAD drawing and SOLIDWORKS model of the complete assembly of the TIRCAM2 mounted on the side port of the DOT.}
\label{fig7}
\end{figure}

Considering the envelop size, mass and focal plane of the side port, the instrument interface plate and rack for mounting the TIRCAM2 were designed as shown in Figures 7, 8 and 9. All camera equipments were suitably mounted within the rack. 

\begin{figure}[!h]
\includegraphics[width=1\columnwidth]{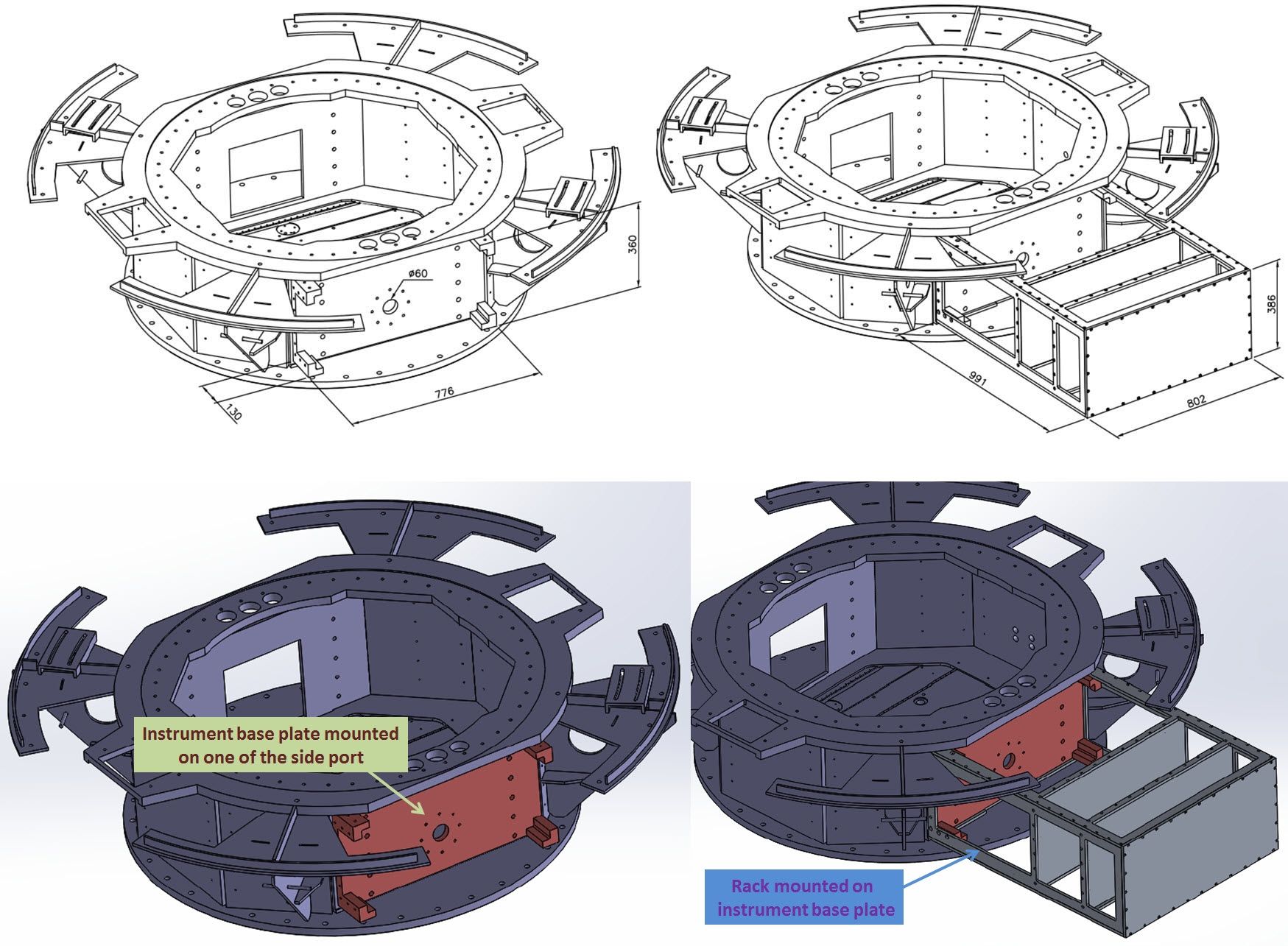}
\caption{AutoCAD drawing and SOLIDWORKS model of the interface plate and rack for the TIRCAM2 mounted on the side port of the DOT Telescope.}
\label{fig8}
\end{figure}

Figure 10 shows instrument base plate (left) and rack (right), with the TIRCAM2 
mounted on the side port and Figure 11 shows the actual photographs of the complete rack assembly with dummy lead bricks for main port (left) and side port (right). Figure 12 shows 
the complete rack assembly with its mounting on the side port of the DOT.
The total weight of the whole assembly is around 248 kg as required for the side port in order to balance the telescope. This 248 kg assembly is lifted by the overhead crane during the mounting on the side port (see Figure 13). UPS supply cables and Ethernet cables are passed through the cable anti-twister in the pier of the telescope. Figure 14 shows the evacuation set up of the TIRCAM2 camera after mounting the camera on the side port. 

\section{Performance at the side port}
During the first light of the TIRCAM2 at the side port, the M53 globular cluster was observed on May 05, 2020. The M53 J-band image was taken at 5 dithered positions with one frame at each position. The image was cleaned for sky-background by subtracting the median of the dithered frames. Mosaic of the M53 is formed by combining the dithered frames. A part of the mosaic (field of view (FoV) $\sim$ $1.45^{\prime}$ $\times$ $1.45^{\prime}$) is shown in Figure 15 (left) and is compared with the corresponding FoV of the 2MASS image (see Figure 15; right). The Palomar 2 globular cluster was observed during regular science observations on November 17, 2021. The J-band image was taken at 5 dithered positions with 11 frames at each position. Image cleaning process was similar to that of the M53 J-band image. A part of the mosaic (FoV $\sim$ $1.3^{\prime}$ $\times$ $1.3^{\prime}$) of the Palomar 2 taken with the TIRCAM2 is shown in Figure 16 (left) and for comparison, the 2MASS J-band image of the same region is also shown in Figure 16 (right). The Full width at half maxiam (FWHM) of the marked sources in the TIRCAM2 images of the M53 (Figure 15; left) and the Palomer 2 (Figure 16; left) are 5.30 pixels ($\sim$ $0.9^{\prime \prime}$) (see Figure 17; left) and 3.53 pixels ($\sim$ $0.6^{\prime \prime}$) (see Figure 17; right) respectively.

\begin{figure}[!h]
\includegraphics[width=1\columnwidth]{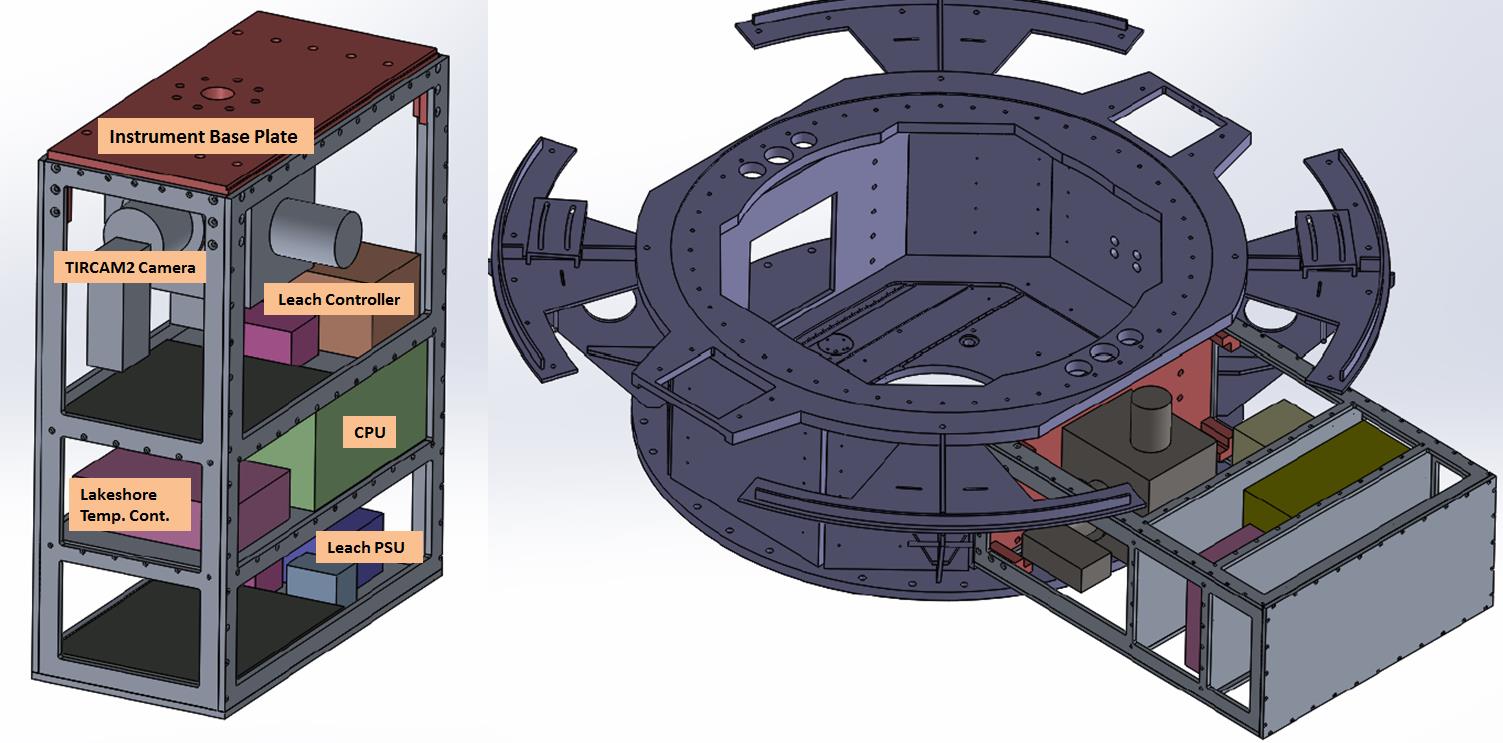}
\caption{AutoCAD drawing and SOLIDWORKS model of the TIRCAM2 rack assembly.}\label{fig9}
\end{figure}

\begin{figure}[!h]
\includegraphics[width=1\columnwidth]{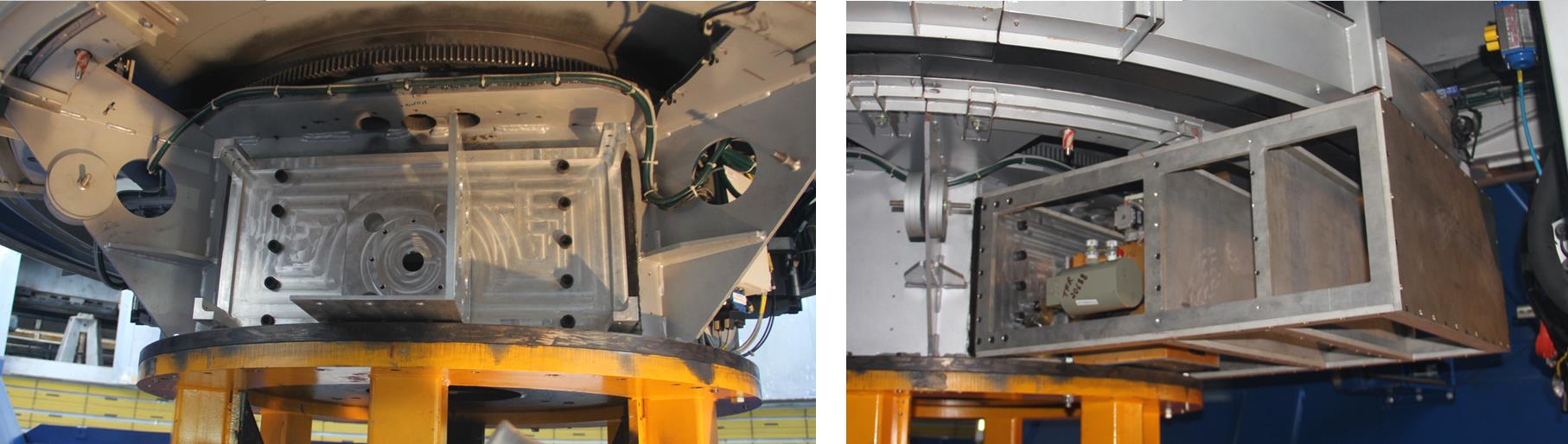}
\caption{Actual photographs of the instrument base plate (left) and only rack (right), with the TIRCAM2 mounted on the side port.}
\label{fig10}
\end{figure}

\section{Achievements at the side port}
With a new optimized mechanical setup, the TIRCAM2 was installed on the sideport of the DOT and it has become the first light instrument on the DOT sideport. A new design and fabrication of the filter mechanism and filter controller were carried out for the sideport by replacing the older filter mechanism and controller. Further tuning of the setup was done especially with the He gas pipe routing within the telescope structure. The TIRCAM2 sideport performance is comparable to the main port performance. The TIRCAM2 is now regularly used for astronomical observations as per scheduled proposals. The TIRCAM2 at the sideport is also regularly used for lunar and planetary occultation observations.

\begin{figure}
\includegraphics[width=1.0\columnwidth]{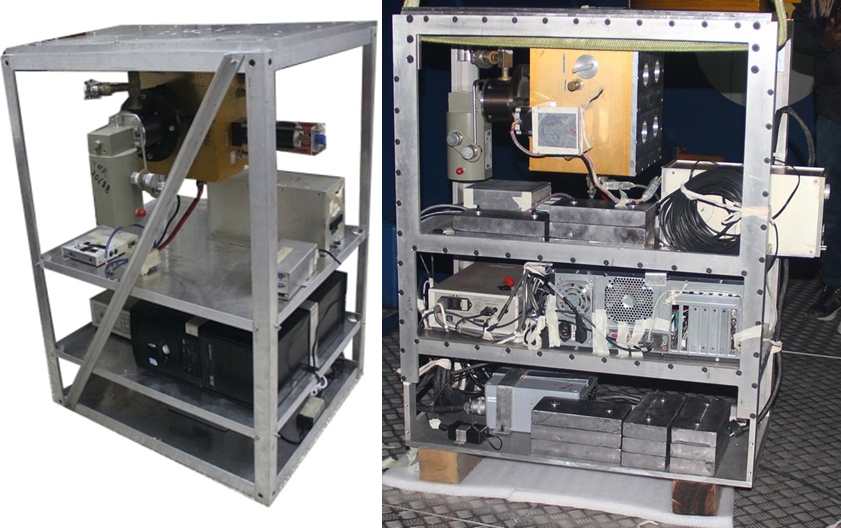}
\caption{The TIRCAM2 complete rack assembly for main port (left) and side port (right) of the DOT.}
\label{fig11}
\end{figure}

\begin{figure}[!h]
\includegraphics[width=1\columnwidth]{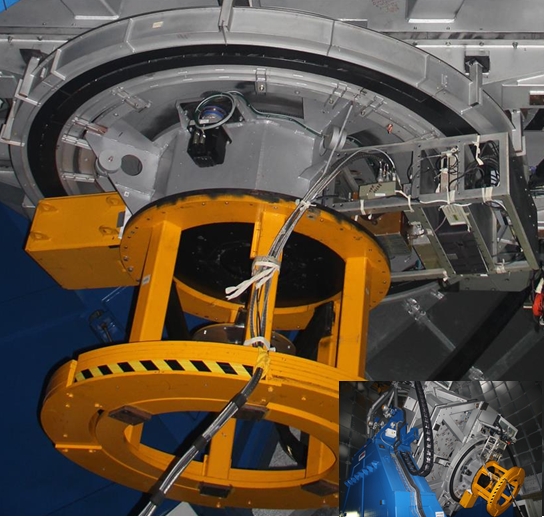}
\caption{The TIRCAM2 instrument rack assembly mounted on the side port of the DOT.}
\label{fig12}
\end{figure}

\begin{figure}[!h]
\includegraphics[width=1\columnwidth]{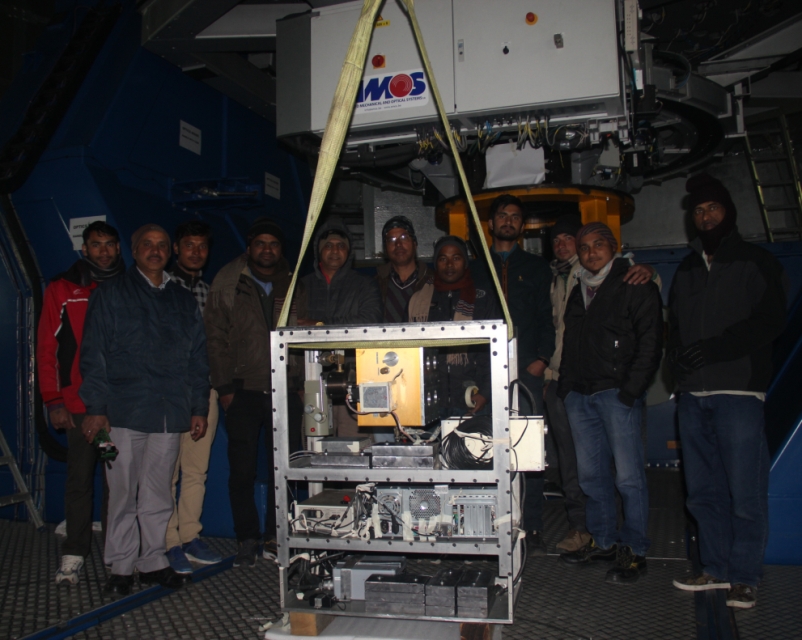}
\caption{The complete 248 kg TIRCAM2 assembly lifted by the overhead crane during the mounting on the side port.}
\label{fig13}
\end{figure}

\begin{figure}[!h]
\includegraphics[width=1\columnwidth]{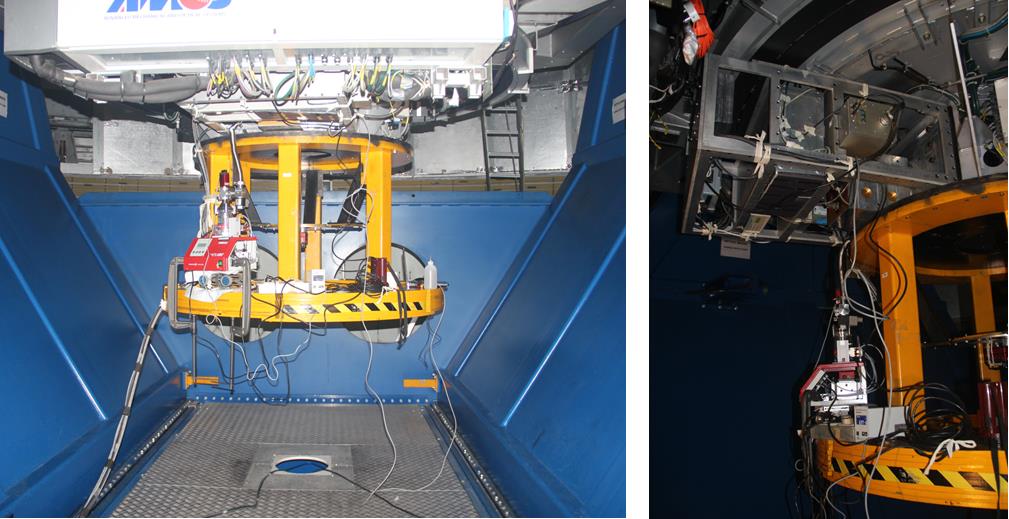}
\caption{The evacuation set up of the TIRCAM2 camera at the telescope floor
after mounting the camera on the side port.}
\label{fig14}
\end{figure}

\begin{figure}[!h]
\includegraphics[width=1\columnwidth]{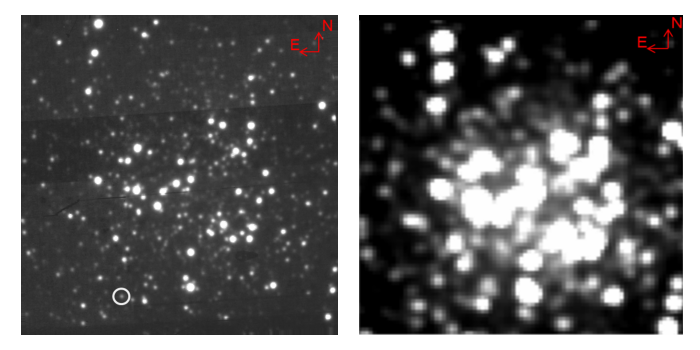}
\caption{(Left) A part of the mosaic ($\sim$ $1.45^{\prime}$ $\times$ $1.45^{\prime}$) image of the M53 globular cluster in the J-band image taken with the TIRCAM2 at the DOT sideport. (Right) The M53 J-band image taken by 2MASS. Both images correspond to the same FoV. North is up, and East is to left}
\label{fig15}
\end{figure}

\begin{figure}[!h]
\includegraphics[width=1\columnwidth]{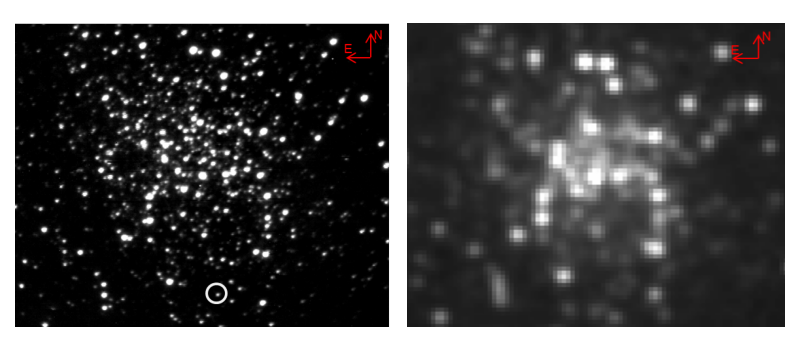}
\caption{(Left) The Palomar 2 globular cluster in the J-band observed with the TIRCAM2 at the sideport. (Right) The Palomar2 J-band image taken by 2MASS in J-band. Both images correspond to the same FoV. North is up, and East is to left.}
\label{fig16}
\end{figure}

\begin{figure}[!h]
\includegraphics[width=1\columnwidth]{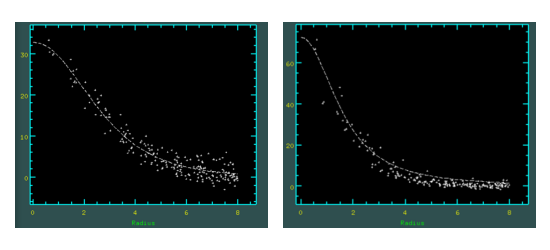}
\caption{The radial profiles of the stars marked in TIRCAM2 J-band images (see Figure 16 and Figure 17) of the M53 (left) and the Palomar 2 globular cluster (right).}
\label{fig17}
\end{figure}

\section{Conclusion}

Interfacing of the TIRCAM2 camera on the side port of the DOT has been completed successfully as shown in Figure 12. The first light with the TIRCAM2 camera set up on the DOT side port was achieved on May 5, 2020. Currently, the TIRCAM2 camera is being used successfully at the DOT side port for science observations.

\section{Acknowledgments}
We thank the anonymous reviewer for several useful suggestions, which greatly improved the technical contents of the paper.
We thank all the staff members of Devasthal Observatory and ARIES workshop for their assistance during installation of the TIRCAM2 camera for the first time on the side port of the telescope. We would also like to thank the staff of TIFR central workshop for fabrication of the TIRCAM2 rack and the instrument interface plate. We also thank Dr. Brijesh Kumar and Dr. T. S. Kumar for their co-operation during the installation of the TIRCAM2 and the balancing of the telescope. S.B.B., M.B.N., S.S.P., H.S., R.B.J., B.J.B., S.L.D., D.K.O. and K.S. acknowledge the support of the Department of Atomic Energy, Government of India, under Project Identification No. RTI 4002.

\bibliography{reference}
\vspace{-2em}

\end{document}